\documentclass[aps,pra,twocolumn,amsmath,showpacs]{revtex4}
\usepackage{bm}
\usepackage{graphicx}
\begin{document}
\setlength{\arraycolsep}{2pt}
\title{Complete conditions for legitimate Wigner distributions}
\author{Hyunchul Nha$^*$}
\affiliation{Department of Physics, Texas A \& M University at Qatar, Doha, Qatar} 
\date{\today}
\begin{abstract}
Given a real-valued phase-space function, it is a nontrivial task to determine whether it corresponds to a Wigner distribution 
for a physically acceptable quantum state. This topic has been of fundamental interest for long, and in a modern application, 
it can be related to the problem of entanglement detection for multi-mode cases. 
In this paper, we present a hierarchy of complete conditions for a physically realizable Wigner distribution. 
Our derivation is based on the normally-ordered expansion, in terms of annihilation and creation operators,
of the quasi-density operator corresponding to the phase-space function in question. 
As a by-product, it is shown that the phase-space distributions with elliptical symmetry 
can be readily diagonalized in our representation, facilitating the test of physical realizability. 
We also illustrate how the current formulation can be connected to the detection of bipartite entanglement 
for continuous variables. 
\end{abstract}
\maketitle
\email{hyunchul.nha@qatar.tamu.edu}

\narrowtext
\section{Introduction}
The phase-space formulation of quantum mechanics has continued to be of great interest 
ever since its first introduction by E.~Wigner \cite{Wigner}. 
Remarkably, J.~E.~Moyal further elaborated on the so-called Wigner-Weyl correspondence 
through an attempt to interpret quantum mechanics as a general statistical dynamics \cite{Moyal}. 
The phase-space treatment of quantum mechanics is particularly interesting in that it provides a valuable insight 
into the issue of quantum-classical correspondence \cite{Hillery0}. It exhibits both the similarities and the differences 
between the quantum and the classical descriptions of physical world. 
Although the original scheme was intended to shed light on quantum mechanics from a classical perspective, 
the inverse problem, i.e., the description of classical mechanics in Hilbert space from a quantum perspective, 
has also attracted much interest \cite{Groenewold,Bracken1,Bracken2,Habib,Muga}.

The most prominent difference between the quantum and the classical theories may be characterized by the uncertainty principle. 
According to it, every classically admissible distribution in phase space is not always allowed by quantum mechanics. 
In classical mechanics, any positive-definite real-valued functions, 
which may be interpreted as joint-probability densities for having definite values of both the position and the momentum, 
are possible in phase space. On the other hand, in quantum mechanics, due to the canonical commutation relation 
$[{\hat x},{\hat p}]=i\hbar$, the uncertainty relation $\Delta x\Delta p\ge\frac{\hbar}{2}$ must be at least 
satisfied by the distributions. 

However, it is also known that the single uncertainty condition $\Delta x\Delta p\ge\frac{\hbar}{2}$ alone 
does not guarantee that the phase-space function corresponds to a quantum mechanically acceptable state 
\cite{Manko,Gosson,Narcowich}.  
In order to more deeply understand how the quantum principles are reflected in phase space, 
it may be of crucial importance to have complete conditions, desirably in analytic form, 
by which to determine whether a given real-valued 
phase-space function may describe a legitimate quantum state or not. 
This actually defines a substantially nontrivial task in contrast to the inverse problem: 
When a certain quantum state described by a density operator $\rho$ is given, 
it is now a well-known, standardized, procedure to obtain the phase-space distributions corresponding 
to various operator orderings \cite{Cahill}. 
However, even the set of classical phase-space distributions that are allowed by quantum mechanics 
is not yet completely identified \cite{Werner}.

In this paper, we derive a hierarchy of complete conditions 
for a legitimate Wigner distribution of $d$ degrees of freedom. 
It is done by using the correspondence between a given real-valued function in phase space and a Hermitian operator $\rho_q$ 
in Hilbert space, which may be called a quasi-density operator \cite{Bracken2}. 
In particular, the quasi-density operator $\rho_q$ is constructed as a power-series expansion of annihilation/creation operators 
in normal-ordering. 
We then require that the quantum fidelity of the quasi-density operator 
with an arbitrary pure state must be non-negative as a sufficient and necessary condition 
for the positivity of $\rho_q$. 
We compare these conditions 
with the Kastler--Loupias--Miracle-Sole (KLM) conditions \cite{Kastler,Loupias}, 
the only systematic conditions previously known to our best knowledge, 
and remark that some practical advantages may arise from our novel approach. As a by-product, we show that 
the quasi-density operator corresponding to a phase-space distribution with elliptical symmetry can be always diagonalized 
in the generalized Fock-state basis, which may facilitate the test of positivity. 
We also illustrate the utility of the derived conditions by examining the Wigner function corresponding 
to a two-mode entangled state under partial transpose \cite{Peres}, 
which may be interpreted as an observable criterion for bipartite entanglement.

This paper is organized as follows. In Sec.~II, some preliminary facts are briefly introduced to be used later, and 
in Sec.~III, the previously known KLM conditions are reviewed in a heuristic manner. 
In Sec.~IV, a set of complete conditions for legitimate Wigner distributions is derived based 
on the normally-ordered expansion of the quasi-density operator, and 
the case of elliptical symmetry in phase space is more specifically considered. 
These results are further extended particularly to two-mode cases in Sec.~V, and 
the formalism is illustrated by an example relating to the problem of detecting bipartite entanglement. 
In Sec.~VI, the results are summarized with concluding remarks.  

Before going to the main part, we need to clarify our usage of operators in this paper. 
The annihilation and the creation operators, $a$ and $a^\dag$, for one degree of freedom satisfy the well-known boson 
commutation relation, $[a,a^\dag]=1$. 
When the annihilation operator is decomposed into real and imaginary parts as $a\equiv {\hat X}+i{\hat Y}$, 
the two Hermitian operators ${\hat X}=\frac{a+a^\dag}{2}$ and ${\hat Y}=\frac{a-a^\dag}{2i}$, 
which are known as the quadrature amplitudes in quantum optics, satisfy $[X,Y]=\frac{i}{2}$. 
In other words, setting $\hbar=1$, the quadrature amplitudes can be related to the position and 
the momentum operators as ${\hat x}=\sqrt{2}{\hat X}$ and ${\hat p}=\sqrt{2}{\hat Y}$. 
In this paper, we will take into consideration only the quadrature operators ${\hat X}$ and ${\hat Y}$
 and their corresponding values in phase space, 
not the canonical operators ${\hat x}$ and ${\hat p}$. 
However, the translation of the results into terms related to ${\hat x}$ and ${\hat p}$ is rather straightforward 
by considering the numerical factor $\sqrt{2}$.

\section{Preliminaries} 

Let us first consider the case with one degree of freedom. It is well known 
that an arbitrary bounded operator $\hat {F}$ 
with a finite Hilbert-Schmidt norm, $||\hat {F}||^2\equiv{\rm Tr} \{\hat {F}^{\dag}\hat {F}\}<\infty$, 
can be represented in an integral form as
\begin{eqnarray}
\hat {F}=\frac{1}{\pi}\int d^2\lambda \hat{D}^\dag(\lambda)C_F(\lambda),
\label{eqn:operator-IR}
\end{eqnarray} 
where $\hat{D}(\lambda)\equiv e^{\lambda a^\dag-\lambda^*a}$ is the displacement operator. 
The complex-valued function 
$C_F(\lambda)={\rm Tr} \{\hat {F}\hat{D}(\lambda)\}$ is usually termed the characteristic function of the operator 
$\hat {F}$ \cite{Cahill}. 
If we further proceed with the normal-ordered expansion of the displacement operator using 
the Baker-Campbell-Hausdorff relation, 
$\hat{D}(\lambda)=e^{\lambda a^\dag-\lambda^*a}=e^{\lambda a^\dag}e^{-\lambda^*a}e^{-\frac{|\lambda|^2}{2}}$, 
the operator $\hat {F}$ can be cast into the form 
\begin{eqnarray}
\hat {F}=\frac{1}{\pi}\sum_{m,n}C_{mn}a^{\dag m}a^n,
\label{eqn:ps}
\end{eqnarray} 
where the coefficients $C_{mn}$ are given by
\begin{eqnarray}
C_{mn}\equiv\frac{(-1)^m}{m!n!}\int d^2\lambda \lambda^m\lambda^{*n}e^{-\frac{|\lambda|^2}{2}}C_F(\lambda). 
\end{eqnarray} 
When the operator $\hat {F}$ is Hermitian, $\hat {F}=\hat {F}^\dag$, then the characteristic function satisfies the property 
$C_F^*(\lambda)=C_F(-\lambda)$, according to which $C_{nm}=C_{mn}^*$ follows. It is also known that the coefficients $C_{mn}$ 
are all finite, bounded as $|C_{mn}|\le\frac{\sqrt{(m+n)!}||\hat {F}||}{m!n!}$, and that the normally-ordered power series 
in Eq.~(\ref{eqn:ps}) converges well to the operator $\hat {F}$ \cite{Cahill}. 

Specifically, when the bounded operator $\hat {F}$ is a quantum density operator $\hat \rho$, 
the characteristic function $C(\lambda)={\rm Tr} \{\hat {\rho}\hat{D}(\lambda)\}$ is used to define the Wigner distribution $W(\alpha)$ 
via a Fourier transform in phase space, i.e., 
\begin{eqnarray}
W(\alpha)=\int d^2\lambda e^{\alpha \lambda^*-\alpha^*\lambda} C(\lambda).
\label{eqn:Wigner}
\end{eqnarray}
The quantum state of any one-dimensional system can be thus represented either by the operator $\hat \rho$ or by the c-number Wigner function $W(\alpha)$, 
as the correspondence in Eq.~(\ref{eqn:Wigner}) is one-to-one. 
For a legitimate quantum state, the density operator $\hat \rho$ is positive semidefinite, i.e., all its eigenvalues are nonnegative, 
along with the trace condition ${\rm Tr}\{\hat {\rho}\}=1$, which is our concern in this paper. 

{\bf legitimate Wigner distribution--} Suppose now that a certain real-valued function $W_t(\alpha)=W_t(\alpha_x,\alpha_y)$ 
is given to be tested whether it corresponds to a legitimate quantum state, or more precisely, a legitimate Wigner distribution. 
One can first take its complex Fourier transform to obtain the characteristic function as 
\begin{eqnarray}
C_t(\lambda)=\int d^2\alpha e^{\lambda \alpha^*-\lambda^*\alpha} W_t(\alpha).
\label{eqn:qd-charact}
\end{eqnarray}
Now, from Eq.~(\ref{eqn:operator-IR}), the quasi-density operator $\rho_q$ 
corresponding to the given $W_t(\alpha)$ is constructed as 
\begin{eqnarray}
\hat {\rho_q}=\frac{1}{\pi}\int d^2\lambda \hat{D}^\dag(\lambda)C_t(\lambda).
\label{eqn:qd-operator}
\end{eqnarray} 
The question at hand is to check if $\hat {\rho_q}$ is positive semidefinite with the trace condition 
${\rm Tr}\{\hat {\rho_q}\}=1$. The trace condition is rather simple to test, 
as ${\rm Tr}\{\hat {\rho_q}\}=C_t(0)=\int d^2\alpha W_t(\alpha)=1$, i.e., it requires that  $W_t(\alpha)$ 
is integrated to unity over the entire phase space. 
Thus, we will focus only on the positive semidefiniteness of $\hat {\rho_q}$ 
throughout this paper.

Note that other types of phase space distributions can be treated in a similar way. Supposed one is given a real-valued function $W_t^s(\alpha)=W_t^s(\alpha_x,\alpha_y)$ 
and asked whether it corresponds to a specific $s$-parametrized distribution in phase space \cite{Cahill}. 
The Wigner distribution corresponds to the case of $s=0$, 
and other important distributions in quantum optics are the Glauber-P function for $s=1$ and the Q-function for $s=-1$. 
One would then calculate the characteristic function as 
\begin{eqnarray}
C_t(\lambda)=e^{-\frac{s}{2}|\lambda|^2}\int d^2\alpha e^{\lambda \alpha^*-\lambda^*\alpha} W_t^s(\alpha),
\label{eqn:qd-charact}
\end{eqnarray} 
to obtain the quasi-density operator in the form of Eq.~(\ref{eqn:qd-operator}). 
In this paper, we will deal only with the case of $s=0$, that is, the Wigner distribution.

\section{KLM conditions} 
To begin with, let us briefly address the KLM conditions 
for the positivity of the quasi-density operator $\hat {\rho_q}$ \cite{nha00}. 
If $\hat {\rho_q}$ is positive semidefinite, the condition ${\rm Tr}\{{\hat f}^\dag{\hat f}\rho_q\}\ge0$ 
must be fulfilled for an arbitrary operator ${\hat f}$. 
Let us particularly take ${\hat f}$ as a discrete sum of the displacement operators, 
i.e., ${\hat f}=\sum_{i=1}^n A_i\hat{D}(\xi_i)$, where $A_i$ and $\xi_i$ are arbitrary complex numbers. 
Then, using the relation
\begin{eqnarray} 
\hat{D}(\alpha)\hat{D}(\beta)=\hat{D}(\alpha+\beta)
e^{\frac{1}{2}(\alpha\beta^*-\alpha^*\beta)}, 
\label{eqn:displacement-sum} 
\end{eqnarray}
and ${\rm Tr}\{\hat{D}(\alpha)\}=\pi\delta^2(\alpha)$, along with Eq.~(\ref{eqn:qd-operator}), it follows 
\begin{eqnarray} 
{\rm Tr}\{{\hat f}^\dag{\hat f}\hat {\rho_q}\}=\frac{1}{\pi}\sum_{i,j=1}^n A_iA_j^*M_{ij}\ge0,
\label{eqn:KLM-inequality}
\end{eqnarray}
where 
\begin{eqnarray}
M_{ij}\equiv e^{\frac{1}{2}(\xi_i\xi_j^*-\xi_i^*\xi_j)}C_t(\xi_i-\xi_j).
\label{eqn:KLM-matrix}
\end{eqnarray}
As the inequality Eq.~(\ref{eqn:KLM-inequality}) must be fulfilled for arbitrary $A_i$'s, 
every $n\times n$ matrix $\{M_{ij}\}$ ($n=1,2,\dots$) must be positive semidefinite, 
which becomes the necessary condition for $\hat {\rho_q}$ to be positive semidefinite. 

According to KLM \cite{Kastler,Loupias}, the converse is also true, i.e., 
the positive semidefiniteness of all matrices $\{M_{ij}\}$ becomes sufficient for that of $\hat {\rho_q}$. 
In our heuristic argument, this sufficiency may be seen, with less mathematical rigor, as follows. 
As implied by Eq.~(\ref{eqn:operator-IR}), an arbitrary operator ${\hat f}$ can be expressed 
as a sum, more precisely an integral, of the displacement operators. 
Then, following the same procedures as before, 
the requirement of ${\rm Tr}\{{\hat f}^\dag{\hat f}\rho_q\}\ge0$ 
for every operator ${\hat f}$ leads to a similar condition to Eq.~(\ref{eqn:KLM-inequality}), 
where the discrete sum is only replaced by a continuous integration.

The KLM conditions may be understood as a quantum version of the Bochner theorem \cite{Bochner,Werner}. 
The "classical" Bochner theorem states that every positive-definite function is a Fourier transform 
of a positive finite Borel measure. 
In our problem, it implies that if the phase-space function is a classical probability density ($W_t(\alpha)\ge0$), 
then the matrix $\{M'_{ij}\equiv C_t(\xi_i-\xi_j)\}$, instead of the one in Eq.~(\ref{eqn:KLM-matrix}), 
must be positive semidefinite. 
Thus, the only difference between quantum and classical cases is 
the additional factor $e^{\frac{1}{2}(\xi_i\xi_j^*-\xi_i^*\xi_j)}$ 
of Eq.~(\ref{eqn:KLM-matrix}) in quantum cases, which obviously arises due to the commutation relation $[a,a^\dag]=1$ 
through Eq.~(\ref{eqn:displacement-sum}).

On constructing the $n\times n$ matrix $\{M_{ij}\}$, one has to show that $M_{ij}\ge0$ for every choice of $\xi_i$ ($i=1,\cdots,n$) 
to confirm the positivity of $\hat {\rho_q}$. 
In many cases, the test may thus amount to the optimization problem involving $2(n-1)$ 
real independent variables corresponding to the complex variables $\xi_i-\xi_{i+1}$ ($i=1,\cdots,n-1$), 
which becomes increasingly hard for a growing number of $n$. 
We will now derive another set of complete conditions for the positivity of $\hat {\rho_q}$ 
that can avoid such a problem. Furthermore, our approach will turn out to make it possible 
to directly evaluate the eigenvalues of $\hat {\rho_q}$ rather easily for the cases of elliptical symmetry in phase space.

\section{Complete conditions for Wigner function} 
Given the phase-space distribution $W_t(\alpha)$, or equivalently, 
the characteristic function $C_t(\lambda)$ in Eq.~(\ref{eqn:qd-charact}), one may represent the quasi-density operator 
in a normally-ordered form given by Eq.~(\ref{eqn:ps}) as 
\begin{eqnarray}
\hat {\rho_q}=\frac{1}{\pi}\sum_{m,n}C_{mn}a^{\dag m}a^n.
\label{eqn:ps-qd}
\end{eqnarray} 
 Here, the coefficients are given by 
\begin{eqnarray}
C_{mn}\equiv\frac{(-1)^m}{m!n!}\int d^2\lambda \lambda^m\lambda^{*n}e^{-\frac{|\lambda|^2}{2}}C_t(\lambda), 
\label{eqn:qd-coeff}
\end{eqnarray}
or, alternatively in terms of the phase-space function $W_t(\alpha)$, by
\begin{eqnarray}
C_{mn}=\frac{2\pi(-2)^m}{m!n!}\int d^2\alpha \left[\alpha^{m-n}S(m,m-n,2|\alpha|^2)W_t(\alpha)\right], 
\label{eqn:qd-coeff1}
\end{eqnarray}
where the kernel $S$ is defined by 
\begin{eqnarray}
S(l_1,l_2,x)&\equiv&\sum_k\frac{(k+l_1)!}{(k+l_2)!k!}(-x)^k\nonumber\\
&=&\frac{l_1!}{l_2!} \hspace{0.001cm}_1F_1[l_1+1,l_2+1,-x].
\label{eqn:hyper}
\end{eqnarray} 
($_1F_1$ refers to the hypergeometric function.)

With the identification of the coefficients $C_{mn}$ for a given distribution $W_t(\alpha)$, 
the next step is to check whether the quasi-density operator in Eq.~(\ref{eqn:ps-qd}) is positive semidefinite. 
This can be done using the fact that a Hermitian operator $\hat {H}$ is positive semidefinite if and only if 
$\langle\Psi|\hat {H}|\Psi\rangle\ge0$ for every pure state $|\Psi\rangle\in H$, 
where $H$ is the Hilbert space under consideration. Expressing an arbitrary pure state in the Fock-state basis as 
$|\Psi\rangle=\Sigma_kD_k|k\rangle$ ($a^\dag a|k\rangle=k|k\rangle$), this condition reads 
\begin{eqnarray}
\langle\Psi|\hat {\rho_q}|\Psi\rangle=\sum_{k,k'}D_k^*D_{k'}\alpha_{kk'}\ge0,
\end{eqnarray}
where 
\begin{eqnarray}
\alpha_{kk'}=\frac{1}{\pi}\sum_{m=M}^k\frac{\sqrt{k!k'!}}{(k-m)!}C_{mk'-k+m},
\label{eqn:alpha}
\end{eqnarray}
($M\equiv {\rm max}\{0,k-k'\}$). 
Therefore, the positive semidefiniteness of the matrix $\{\alpha_{kk'}\}$ becomes 
the sufficient and necessary condition for the legitimate Wigner distributions. 

As a matter of fact, $\alpha_{kk'}$ is simply the matrix element of the quasi-density operator in the Fock state basis, 
$\alpha_{kk'}=\langle k|\hat {\rho_q}|k'\rangle$. In this regard, if one directly calculate $\alpha_{kk'}$ from 
Eq.~(\ref{eqn:qd-operator}) 
using the matrix element of the displacement operator, 
$\langle k|\hat{D}^\dag(\lambda)|k'\rangle=\sqrt{\frac{k'!}{k!}}
(-\lambda)^{k-k'}e^{-|\lambda|^2/2}L_{k'}^{(k-k')}(|\lambda|^2)$, 
where  $L_q^{(p)}(x)$ is an associated Laguerre polynomial \cite{Cahill}, 
it follows 
\begin{eqnarray}
\alpha_{kk'}=\frac{(-1)^{k-k'}}{\pi}\sqrt{\frac{k'!}{k!}}
\int d^2\lambda\left[\lambda^{k-k'}e^{-\frac{|\lambda|^2}{2}}C_t(\lambda)L_{k'}^{(k-k')}(|\lambda|^2)\right],
\label{eqn:alphaa}
\end{eqnarray}
In this paper, instead of Eq.~(\ref{eqn:alphaa}), we are going to deal with the expression in Eq.~(\ref{eqn:alpha}), 
which is more directly associated with the normal-ordered operator form, Eq.~(\ref{eqn:ps-qd}),  
by way of the coefficients $C_{mn}$. Note that the condition $\{\alpha_{kk'}\}\ge0$ has a clear physical interpretation 
as the positivity of quantum fidelity of the quasi-density operator with arbitrary pure states. 
This is a full generalization of the approach taken by Manko {\it et al.} for a particular example in \cite{Manko}.

Clearly, the condition $\{\alpha_{kk'}\}\ge0$ does not involve the optimization problem 
in contrast to the KLM conditions. 
Instead, all relevant information is incorporated 
by evaluating the coefficients $C_{mn}$ of the quasi-density operator $\hat {\rho_q}$ in the form of Eq.~(\ref{eqn:qd-coeff}) 
or Eq.~(\ref{eqn:qd-coeff1}). 
The positive semidefiniteness of the matrix $\{\alpha_{kk'}\}$ can be characterized only 
in terms of matrix determinants by Sylvester's criterion \cite{Horn}. That is, 
all principal minors constructed from the matrix $\{\alpha_{kk'}\}$ must be nonnegative 
for a given distribution $W_t(\alpha)$ to represent a physically realizable quantum state.

{\bf Case of elliptical symmetry--} Let us now consider the case in which the distribution function $W_t(\alpha_x,\alpha_y)$ 
possesses elliptical symmetry in phase space, more precisely, the value of $W_t$ depends only on the parameter
$r_\alpha\equiv \sqrt{\frac{(\alpha_x'-\beta_x)^2}{a^2}+\frac{(\alpha_y'-\beta_y)^2}{b^2}}$, where 
$\alpha_x'\equiv \alpha_x\cos\phi-\alpha_y\sin\phi$, and $\alpha_y'\equiv \alpha_x\sin\phi+\alpha_y\cos\phi$. 
($a,b$: real constants)
In other words, the distribution is centered at the point $(\beta_x,\beta_y)$, and the major/minor axes of the ellipse 
are rotated by an angle $\phi$. The general elliptical distribution can be transformed to a standard one centered at origin 
with $x$- and $y$-axis as major/minor axes by a unitary operation as $U\hat {\rho_q}U^\dag$. Here, $U$ is 
the displacement operator followed by the phase-shift, $U\equiv e^{-i\phi a^\dag a}D(-\beta)$. 
As the unitary transformation does not change the positivity of the Hermitian operator, 
one may focus on the positivity of the elliptic distribution only in a form of $W_t(\alpha_x,\alpha_y)=
W_t(\sqrt{\frac{\alpha_x^2}{a^2}+\frac{\alpha_y^2}{b^2}})$ without loss of generality.   

Furthermore, let us now consider a rescaled bosonic operator as $a'\equiv\sqrt{\frac{a}{b}}{\hat X}+i\sqrt{\frac{b}{a}}{\hat Y}$, 
where ${\hat X}$ and ${\hat Y}$ are the original quadrature operators \cite{nhaa}. 
The new operator $a'$ and its conjugate $a'^\dag$ obviously define a bosonic mode 
as the commutation relation $[a',a'^\dag]=1$ holds. The corresponding rescaled distribution then becomes 
$W_t(\alpha_x,\alpha_y)=W_t(\frac{1}{\sqrt{ab}}|\alpha|)$, i.e., it now possesses the circular symmetry in phase space. 
In this case, the quasi-density operator becomes diagonal in the Fock-state basis, 
as the coefficients $C_{mn}=0$ for $m\ne n$ in Eq.~(\ref{eqn:qd-coeff}) 
or in Eq.~(\ref{eqn:qd-coeff1}). ( Note that $[a^{\dag m}a^m,a^{\dag}a]=0$ for every integer $m$.)
When $\hat {\rho_q}$ is expressed as $\hat {\rho_q}=\frac{1}{\pi}\sum_{m}C_{mm}a^{\dag m}a^m$, the diagonal terms become 
$\alpha_{kk}=\langle k|\hat {\rho_q}|k\rangle=\frac{1}{\pi}\sum_{m=0}^k\frac{k!}{(k-m)!}C_{mm}$, 
which of course correspond to the eigenvalues of $\hat {\rho_q}$. 
The test of positivity of $\hat {\rho_q}$ thus becomes relatively easy for the elliptical distributions using our method.

As an example, let us consider the distribution introduced by Manko {\it et al.} in \cite{Manko}, 
which is obtained by rescaling the initial Wigner distribution 
as $\alpha_x\rightarrow\lambda_x\alpha_x$ and $\alpha_y\rightarrow\lambda_y\alpha_y$. 
Manko {\it et al}. particularly showed that for $\lambda_x=\lambda_y=\lambda$, 
the uncertainty relation $\Delta x\Delta p\ge\frac{\hbar}{2}$ is fulfilled with the condition $\lambda\le1$, 
and that the original Wigner distribution for the Fock state $|1\rangle$, however, becomes unphysical for very small $\lambda$.

In fact, using our method, one can easily check that the rescaled Wigner distributions become unphysical 
for any values of $\lambda\ne1$. 
The deformed characteristic function for $|1\rangle$ reads
\begin{eqnarray}
C(\xi)=\left(1-\left|\frac{\xi}{\lambda}\right|^2\right)e^{-\frac{1}{2}\left|\frac{\xi}{\lambda}\right|^2},
\end{eqnarray}
where $\lambda\equiv\sqrt{\lambda_x\lambda_y}$. Then, the coefficients in Eq.~(\ref{eqn:qd-coeff}) are obtained as
\begin{eqnarray} 
C_{mm}=\frac{(-1)^m\pi}{m!(1+\lambda^2)}\left(\frac{2\lambda^2}{1+\lambda^2}\right)^{m+1}(\lambda^2-1-2m), 
\end{eqnarray} 
and the eigenvalues of the quasidensity operator given by 
\begin{eqnarray}
\alpha_{kk}=\frac{2\lambda^2}{(1+\lambda^2)^3} \left(\frac{1-\lambda^2}{1+\lambda^2}\right)^{k-1}
\left[4k\lambda^2-(1-\lambda^2)^2\right],
\end{eqnarray}
where $k=0,1,2,\cdots$. With $\lambda<1$, the eigenvalue $\alpha_{00}$ always becomes negative. 
On the other hand, with $\lambda>1$, the eigenvalues $\alpha_{kk}$ always become negative 
for $k>\frac{(\lambda^2-1)^2}{4\lambda^2}$. 
Therefore, only the trivial rescaling, $\lambda=1$, preserves the physical realizability of the Wigner distributions.

\section{Multi-mode cases} 

It is rather straightforward to extend the previous results to cases of $d$ degrees of freedom. 
Specifically, for two-mode cases, the quasi-density operator may be represented as 
\begin{eqnarray}
\hat {\rho_q}=\frac{1}{\pi^2}\sum_{{\bf m}{\bf n}}C_{m_1n_1m_2n_2}a_1^{\dag m_1}a_1^{n_1}a_2^{\dag m_2}a_2^{n_2}.
\label{eqn:ps-qd-two}
\end{eqnarray} 
where
\begin{eqnarray}
C_{m_1n_1m_2n_2}\equiv\prod_{i=1,2}\left[\frac{(-1)^{m_i}}{m_i!n_i!}\right]
\int \prod_{i=1,2}\left[d^2\lambda_i\lambda_i^{m_i}\lambda_i^{*n_i}
e^{-\frac{1}{2}|\lambda_i|^2}\right]C_t(\lambda_1,\lambda_2). 
\label{eqn:qd-coeff-two}
\end{eqnarray} 
The characteristic function $C_t(\lambda_1,\lambda_2)$ in Eq.~(\ref{eqn:qd-coeff-two}) can be obtained 
in terms of the two-mode phase-space distribution $W_t(\alpha_1,\alpha_2)$ as 
\begin{eqnarray}
C_t(\lambda_1,\lambda_2)=\int \prod_{i=1,2} \left[d^2\alpha_i e^{\lambda_i \alpha_i^*-\lambda_i^*\alpha_i}\right] 
W_t(\alpha_1,\alpha_2),
\label{eqn:qd-charact-two}
\end{eqnarray}
which leads to an alternative expression for $C_{m_1n_1m_2n_2}$ as 
\begin{eqnarray}
C_{m_1n_1m_2n_2}=\prod_{i=1,2}\left[\frac{2\pi(-2)^{m_i}}{m_i!n_i!}\right]
\int \prod_{i=1,2}\left[d^2\alpha_i \alpha_i^{m_i-n_i}
S(m_i,m_i-n_i,2|\alpha_i|^2)\right] W_t(\alpha_1,\alpha_2),\nonumber\\&& 
\label{eqn:qd-coeff1-two}
\end{eqnarray}
where the kernel $S$ is defined in Eq.~(\ref{eqn:hyper}). 
Taking similar steps, one obtains the condition for the positive semidefiniteness of the quasi-density operator as that of 
the matrix $\{\alpha_{{\bf k}{\bf k'}}\}$, where 
\begin{align}
\alpha_{{\bf k}{\bf k'}}=\frac{1}{\pi^2}\sum_{m_1=M_1}^{k_1}\sum_{m_2=M_2}^{k_2}
\left[\frac{\sqrt{k_1!k_1'!k_2!k_2'!}}{(k_1-m_1)!(k_2-m_2)!}C_{m_1,k_1'-k_1+m_1,m_2,k_2'-k_2+m_2}\right].
\label{eqn:mat-elem-two}
\end{align}
($M_i\equiv {\rm max}\{0,k_i-k_i'\}, i=1,2$). 
Note that we have used the collective indices ${\bf k}\equiv \{k_1,k_2\}$ and ${\bf k'}\equiv \{k_1',k_2'\}$ 
in Eq.~(\ref{eqn:mat-elem-two}). 
In constructing the matrix $\{\alpha_{{\bf k}{\bf k'}}\}$, 
these indices may be arranged in the increasings order of the sums $k=k_1+k_2$ and $k'=k_1'+k_2'$, respectively. 
When the sum is fixed, e.g., $k=k_1+k_2$, one takes the order as $\{k_1,k_2\}=\{0,k\},\{1,k-1\},\cdots,\{k,0\}$, 
and similarly for $\{k_1',k_2'\}$.

{\bf Detection of bipartite entanglement--}
As an illustration, let us consider the problem of detecting bipartite entanglement for continuous variables (CVs). 
The entanglement criteria so far known for CVs are all based on the partial transposition (PT) 
\cite{Simon,Shchukin,Hillery,Agarwal,nha1}. 
When a two-mode state $\rho$ is separable, it is represented by a form 
$\rho=\sum_ip_i\rho_1^{(i)}\otimes\rho_2^{(i)}$, 
where the state $\rho_j^{(i)}$ refers to the subsystem $j=1,2$. Under partial transposition (PT) for the subsystem 2, 
a separable state still remains positive, therefore it describes a certain physical state \cite{Peres}. 

In our formalism, for a given two-mode entangled state with the Wigner distribution $W_t(\alpha_1,\alpha_2)$, 
one can take the partially transposed distribution, 
$W_t(\alpha_1,\alpha_2^*)$ \cite{Simon,nha1}. 
If the state becomes negative under PT, 
then the distribution $W_t(\alpha_1,\alpha_2^*)$ must violate at least one of the nonnegative conditions 
for the matrix $\{\alpha_{{\bf k}{\bf k'}}\}$, which demonstrates the presence of entanglement. 
Note that the coefficients $C_{m_1n_1m_2n_2}^{\rm PT}$ 
for the partially transposed distribution are given by the original coefficients as 
$C_{m_1n_1m_2n_2}^{\rm PT}=C_{m_1n_1n_2m_2}$ through Eqs.~(\ref{eqn:qd-coeff-two}) and~(\ref{eqn:qd-charact-two}).

For instance, let us consider the two-mode entangled state $|\Psi\rangle=\alpha|00\rangle+\beta|11\rangle$. 
The determinant of the $3\times3$ matrix $\{\alpha_{{\bf k}{\bf k'}}\}$ for the partial-transpose distribution is 
obtained in terms of the original coefficients $C_{m_1n_1m_2n_2}$ for the state $|\Psi\rangle$ as  
\begin{eqnarray}
\frac{1}{\pi^2}&&\left|\begin{pmatrix}
&C_{0000}&C_{0010}&C_{0100}\\&C_{0001}&C_{0000}+C_{0011}&C_{0101}\\&C_{1000}&C_{1010}&C_{0000}+C_{1100}
\end{pmatrix}\right|\nonumber\\
&&=\left|\begin{pmatrix}
&|\alpha|^2&0&0\\&0&0&\alpha\beta^*\\&0&\alpha^*\beta&0
\end{pmatrix}\right|<0,
\label{eqn:entangle}
\end{eqnarray}
which is negative for any nonzero values of $\alpha$ and $\beta$. 

The above inequality in Eq.~(\ref{eqn:entangle}) implies that by measuring several coefficients of $C_{m_1n_1n_2m_2}$ 
for a given two-mode state, one can verify bipartite entanglement. To experimentally obtain those values, however, 
it seems that one must construct first the two-mode Wigner distribution via the homodyne tomography \cite{Leonhardt}, 
as seen from Eq.~(\ref{eqn:qd-coeff1-two}). This may be regarded as a practical disadvantage, 
although one could not rule out the possibility to construct the coefficients through an alternative, 
desirably more efficient, experimental method. On the other hand, there is also a certain advantage 
in our formulation regarding entanglement detection. Since we formulated the complete conditions for Wigner distributions 
from the positivity of the quantum fidelity of the quasi-density operator with respect to pure states in the Fock-state basis, 
one can effectively constrain the dimension of the matrix $\{\alpha_{{\bf k}{\bf k'}}\}$ to take into account. 
Specifically, when the total excitation of the state is bounded by $N$, one just need to construct at most $N'\times N'$ matrix, 
where $N'\equiv\frac{(N+1)(N+2)}{2}$. This was actually done for the above example of 
$|\Psi\rangle=\alpha|00\rangle+\beta|11\rangle$, for which a further reduction of the dimension of the matrix was possible 
in Eq.~(\ref{eqn:entangle}). 

\section{Summary and Concluding Remarks}
In summary, we have derived a set of complete conditions for a legitimate Wigner distribution of $d$ degrees of freedom, 
based on the normally-ordered expansion of the quasi-density operator in terms of annihilation/creation operators. 
It was argued that this set may provide a practical advantage over the previously known KLM conditions. 
Furthermore, it was shown that the phase-space distributions with elliptical symmetry can be rather easily 
diagonalized in our formalism, thereby facilitating the test of positivity of the quasi-density operator. 
The derivation was specifically extended to two-mode cases and it was illustrated how the conditions can be 
used in detecting bipartite entanglement for CVs, along with the discussion on its experimental implementation.

In this paper, we formulated the complete conditions for Wigner distributions using the positivity of quantum fidelity of 
the quasi-density operator with arbitrary pure states particularly in the Fock-state basis $|n\rangle$. 
In one perspective, it would be worthwhile to pursue this quantum-fidelity-based approach using generalized basis states $|b_n\rangle$ ($\sum_n|b_n\rangle\langle b_n|=I$), 
particularly to identify other types of phase-space symmetry for which a diagonalization rather easily follows. 
In another perspective, it was recently proved that the positive semidefiniteness of a Hermitian operator is equivalent to the satisfaction of 
{\it all} uncertainty relations \cite{nha}. It was further shown 
that a separability inequality can be systematically derived for any given negative PT entangled state. 
In this respect, it also seems worthwhile to derive another equivalent set of conditions 
for admissible Wigner distributions purely in terms of uncertainty relations in phase space, which is left for future work.

The author greatly acknowledges H.~J.~Carmichael for helpful discussions and the hospitality during his visit 
to the University of Auckland where the current work was carried out. 
This work is supported by a grant from the Qatar National Research Fund.  

*email: hyunchul.nha@qatar.tamu.edu


\begin{references}
\bibitem{nha00} The derivation of the KLM conditions in this paper is rather heuristic and adapted to 
the formulation by Cahill and Glauber \cite{Cahill}.
\bibitem{nhaa} A similar transformation was also used for two specific examples in Ref.~\cite{Bracken1}.
\end{references}
\end{document}